# Oxygen atom density and kinetics in intermediate-pressure radiofrequency capacitively-coupled plasmas in pure $O_2$

Shu Zhang, Andrey Volynets, Garrett A. Curley and Jean-Paul Booth

**Abstract**

We have studied radiofrequency (13.56 MHz) capacitively-coupled plasmas in pure $O_2$ using single-mode laser cavity ringdown spectroscopy of oxygen atoms at 630 nm. The absolute atom densities and translational temperatures were determined over a range of pressures (67-800 Pa) and RF power (50 – 900 W). At pressures of 267 Pa and above, the O-atom mole-fraction increases with RF power and decreases with pressure, reaching a maximum of 15%. However, at 133 and 67 Pa it passes through a distinct maximum with power before decreasing significantly. The atom recombination processes are probed by time-resolved measurements in the afterglow of pulse-modulated plasmas. At 133 and 67 Pa the atom loss is dominated by surface recombination, and we see clear evidence that this rate is increased by energetic ion bombardment, in agreement with a study from Bill Graham's group [1]. This effect partially explains the observed decrease in dissociation at high RF power. The time-resolved results also allow the $O^-$ negative ion density to be determined and indicate the creation of ozone in the afterglow. At 133 Pa, the trends with RF power of the $O_2$ dissociation, $O^-$ density and gas temperature suggest a transition at high power to a plasma mode with fewer high-energy electrons. At higher pressures gas phase recombination mechanisms become dominant, however gas convection driven by gas cooling in the afterglow makes it complex to analyse the time-resolved data.

# 1. Introduction

Low-temperature plasmas containing oxygen gas are widely used in many plasma processing applications, including polymer surface modification [2,3], medicine [4], cleaning of various surfaces [5,6], sterilization of medical devices [7,8], plasma-based water treatment [9], etching of various materials [10,11]. Notably, radiofrequency capacitively-coupled plasmas (RF CCPs) containing oxygen gas at intermediate pressure are used for chemical vapor deposition of oxide films [12,13] and for resist stripping in microelectronics [14,15], due to the high density of O atoms generated in such plasmas. The most important parameters characterizing this system are the electron density and electron energy distribution function (EEDF), which will be discussed in an accompanying paper, and the density of reactive neutral species, especially oxygen atoms.

There have been several experimental studies of the oxygen atom densities in $O_2$ RF CCPs. Kechkar *et al.* [16] studied oxygen RF CCP over a pressure range of 13.3 Pa to 53.2 Pa, using two-photon laser-induced fluorescence (TALIF) to investigate the influence of $SF_6$ on the O atom density. Katsch *et al.* compared the O atom measurement by actinometry and by TALIF in the Gaseous Electronics

Conference reference cell [17]. They also investigated the axial distributions of negative oxygen ions and their power dependency [18]. Takeshi *et al.* [19] studied the influence of the driving frequency on the absolute oxygen atom density in pulse-modulated $O_2$ RF CCP using vacuum ultraviolet absorption spectroscopy. However, previous works are mostly between 1 to 50 Pa, matching only our lowest pressure. Higher-pressure RF CCPs in oxygen, which are used for thin-film deposition, have been little studied apart from one paper by Lisovskiy *et al.* [20]. Therefore, a comprehensive data set over a range of oxygen pressures, and including measurements of both the electrical characteristics (voltage, current and phase, electron density, ion fluxes) and the neutral composition and temperature, is needed in order to test and validate models of this system. The latter dataset, neutral composition and temperature, is the subject of this paper.

Various diagnostic methods have been developed to measure oxygen atom density. Non-optical methods, including chemical titration using NO molecules [21–23], threshold-ionization mass-spectrometry [24] and Catalytic Probes [25,26] have been developed, but they are invasive. Optical methods are preferable, as they do not perturb the plasmas and allow localized density measurements with high temporal resolution. The simplest technique to apply (although rigorous interpretation can be complex), is optical emission spectroscopy (OES) using the emission at 844 and 777 nm [17,27–30]. Absolute calibration (actinometry) can be achieved by calibrating against the OES from actinometer gases (Ar or Xe) added in small quantity, but the accuracy depends on knowledge of the cross-section near the threshold and the shape of the EEDF[31].

TALIF is widely-used because of its high sensitivity and excellent temporal and spatial resolution. An absolute calibration scheme using TALIF of Xe was developed by Niemi *et al.* [32], although the accuracy of this calibration has been disputed recently by Drag *et al.* [33]. Furthermore, the highly non-linear nature of this technique leads to mediocre signal-to-noise ratio.

Absorption-based methods are preferable for absolute density measurements, since they are linear, and do not require any calibration. A disadvantage is that they only give line-integrated measurements, unless optical access allows for an Abel inversion of an axially-symmetric target. Resonance absorption in the vacuum ultraviolet (130 nm) has long been used for measuring O atom density [34–36]. However, the very large absorption cross-section of this allowed transition causes the absorption to saturate even at low column densities (around $10^{14}$ cm$^{-3}$), and therefore it is not applicable to the conditions of this study.

Peverall *et al* [37] developed oxygen atom measurement using single-mode diode laser Cavity Ring Down Spectroscopy (CRDS) of the forbidden $O(^3P_2)$ to $O(^1D_2)$ transition at 630 nm. Even with cavity enhancement, this technique is only applicable to relatively high column densities (above about $10^{15}$ cm$^{-3}$) but can be perfectly adapted to intermediate-pressure discharges in $O_2$ [38].

Here we present an experimental study of RF CCP $O_2$ discharges at intermediate pressure ranges (67-400 Pa) using 630 nm CRDS. The trends with pressure and RF power are presented, as well as the time-resolved densities in pulse-modulated plasmas, giving qualitative insights into the atom creation and loss processes. Complete modelling of the charged particle and neutral dynamics is necessary to fully understand these results; this study provides the quantitative data needed to test and validate such models as they become available.

## 2. Experimental Setup

### 2.1 Plasma Chamber

A schematic of the RF-CCP chamber, named "Dracula" (**D**iagnostics of **RA**diofrequency **C**CP in Large **A**rea), is shown in **Fig. 1**

**Fig. 1**. The discharge is created between two 50 cm diameter aluminium electrodes, separated by 2.5 cm. The upper electrode is grounded, and radiofrequency power at 13.56 MHz is coupled to the lower electrode. Note that this is a highly symmetric system, with a dielectric radial boundary (composed of a glass ring facing the plasma, surrounded by a thick PTFE ring). Pure oxygen gas is flowed into the reactor via the tubes connecting the CRDS mirrors (to protect them from the plasma and reactive radicals). The gas is pumped out through radial slots in the upper electrode and evacuated through the upper manifold by a primary pump with a manual throttle valve to set the chamber pressure.

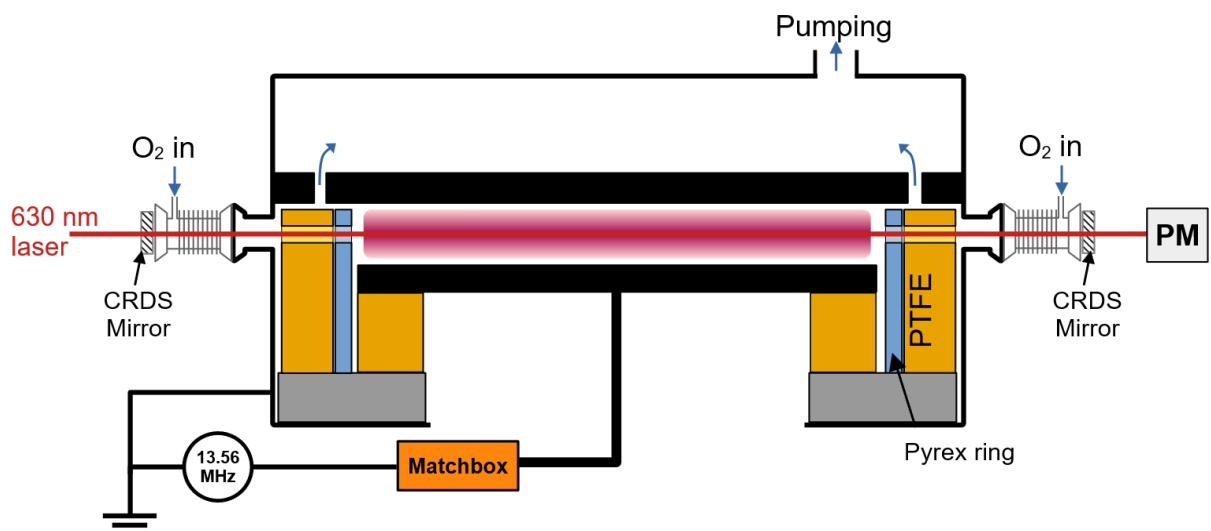

**Fig. 1.** RF CCP chamber

Radiofrequency power is supplied by a 1500 W generator operating at 13.56 MHz (Comet Synertia® RFG 15/13). Impedance matching is achieved using a manual L-type matchbox with two tuneable capacitors $C_{Load}$ (150 – 1500 pF) and $C_{Tune}$ (10 – 500 pF), and a fixed inductor of 2 μH. Pulsed operation of an RF-CCP in this pressure range can be problematic due to difficulty with striking the plasma. This is because the electrical impedance of the chamber without plasma is significantly different form the case with a steady-state plasma: if the impedance matching network is optimised for steady-state operation, the electrode voltage is insufficient to initiate breakdown. To overcome this problem, the RF generator uses variable frequency, allowing resonance to be achieved before breakdown, providing the necessary overvoltage. The power delivered to the match box is measured both by the power meter of the RF generator and by an in-line power meter (Rohde and Schwarz NRT with NAP-Z8 detector head), situated between the generator and the match box. An RF impedance probe (Impedans Octiv suite) is installed between the match box and the chamber, monitoring the voltage, current and phase supplied to the chamber.

## 2.2 CRDS measurements

The oxygen ground state ($^3P_2$) atom line-integrated density and translational temperature is determined by mono-mode laser CRDS of the spin-forbidden transition O($^3P_2$) → O($^1D_2$) at 630.205 nm (15867.862 cm$^{-1}$), as first demonstrated by Peverall *et al.* [37]. Measurements were made in both continuous steady-state discharges, and in pulse-modulated discharges with time-resolution. The latter measurements, in addition to providing information on the oxygen atom kinetics, also provide measurements of the density of O$^-$ negative ions in the active plasma, and O$_3$ generation in the afterglow from the continuum absorption underlying the oxygen atom peak.

### 2.2.1 Optical Setup for CRDS

The optical setup for the CRDS measurement is presented in **Fig. 2** and is very similar to that used by Booth *et al.* [38] in a DC glow discharge. Two spherical high-reflectivity mirrors (Layertek, 1 m radius, reflectivity ~0.99994 at 630 nm) are placed at opposite sides of the plasma chamber, separated by $L_{cavity} \approx 92$ cm. The characteristic ringdown time of an empty cavity is ~ 50 μs, corresponding to mirror losses of about $6 \times 10^{-5}$. The mirrors are mounted in kinematic mounts and connected to the chamber via stainless-steel bellows. The laser beam is connected to the discharge zone ($L = 50$ cm) via 10 mm holes in the dielectrics, fitted with glass capillary tubes (internal diameter 3 mm), which were necessary to prevent breakdown occurring in these access ports.

A tunable laser beam around 630 nm (~15867.862 cm$^{-1}$, ±0.15 cm$^{-1}$) is generated by a continuous single-mode external cavity tunable diode laser (Toptica DLPro). The wavenumber of the laser output is continuously monitored by a wavemeter (High Finesse WS8, resolution 0.001 cm$^{-1}$). The laser beam passes through an opto-acoustic modulator (AOM, AA OptoElectronic MT110-A1-VIS), and the first-order scattered beam is coupled into an optical fibre. The output from the fibre (~100 μW) is focussed by an adjustable collimator (Schäfter+Kirchhoff 60FC-4-M8-10) to match the curvature of the wavefront to that of the cavity mirrors, ensuring optimal coupling into the TEM(0,0) transverse mode of the CRDS cavity. The radiation leaving the exit mirror passes through a 630 nm interference filter and is detected by a photomultiplier (PMT, Hamamatsu R928). The experiment timing and data acquisition is controlled through a National Instruments (NI) USB 6356 interface card, controlled by a LabVIEW program running on a PC. The PMT signal is digitized by the A/D converter (1.25 MS s$^{-1}$) on the NI interface card.

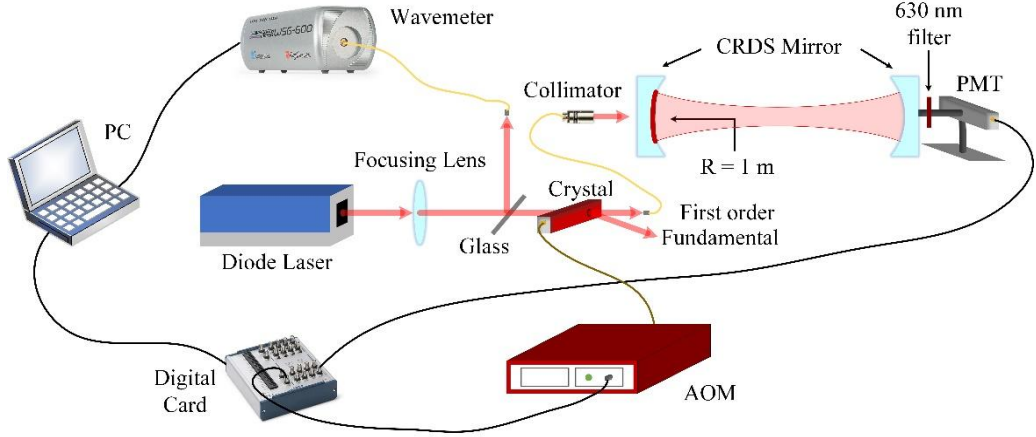

**Fig. 2.** Optical Setup for CRDS measurements.

The laser spectral width (~$10^{-6}$ cm$^{-1}$, but with a jitter of ~$10^{-3}$ cm$^{-1}$) is much narrower than the longitudinal mode spacing of the cavity (~0.005 cm$^{-1}$.). Consequently, it is necessary to achieve longitudinal mode matching of the cavity, which is accomplished by scanning the cavity length at a frequency of several 10's Hz using a piezo actuator attached to the exit mirror. Whenever the cavity length is equal to an integral number of wavelengths of the laser, longitudinal mode matching is achieved, and light builds up in the cavity. This causes a strong increase in the radiation exiting the cavity, detected by the PMT. When the PMT signal intensity passes a set threshold, the AOM is shut off, stopping light injection into the cavity, and the light in the cavity decays exponentially with a "ringdown time", $\tau$.

The linear absorption coefficient, $\alpha_\nu$, (conventionally expressed in units of cm$^{-1}$, averaged over the cavity length) at a given laser wavenumber, $\nu$, due to the presence of an absorbing species (such as O($^3$P$_2$) in this case) can then be calculated from the wavenumber-dependent ringdown time, $\tau_\nu$, as follows:

$$\alpha_\nu = \frac{1}{c\tau_\nu} - \frac{1}{c\tau_0} \qquad \textbf{Eq. 1}$$

where $c$ is the light speed in vacuum, and $\tau_0$ is the ringdown time in the absence of any absorbing species in the cavity. The linear absorption coefficient is directly related to the (absorption-path averaged) number density, $\overline{[X]}$, of an absorbing species X, by:

$$\alpha_\nu = \overline{[X]} \cdot \sigma_\nu^X \qquad \textbf{Eq. 2}$$

where $\sigma_\nu^X$ is the absorption cross-section of species X at wavenumber $\nu$. For an atomic species, it is convenient to integrate this over the line profile (dominated by Doppler broadening in this case) to obtain:

$$\overline{[X]} = \frac{\int \alpha_\nu d\nu}{\int \sigma_\nu^X d\nu} \qquad \textbf{Eq. 3}$$

where the integrated absorption cross-section, $\int \sigma_\nu^X d\nu$, is related to the Einstein A coefficient for the transition in emission, $A_{low}^{up}$, by:

$$\int \sigma_\nu^X d\nu = \frac{g_{up}}{g_{low}} \frac{A_{low}^{up} \lambda^2}{8\pi c} \qquad \text{Eq. 4}$$

where $g_{up}$ and $g_{low}$ are the degeneracies of the upper and lower states and taking care to express all parameters in the same length units (cm or m depending on the case).

### 2.2.2 Measurements of O Atom Density and Gas Temperature in continuous discharges

For continuous plasma measurements, the laser is scanned over the Doppler profile of the absorption line $O(^3P_2) \rightarrow O(^1D_2)$, to obtain the linear absorption coefficient $\alpha_\nu$, as a function of the laser wavenumber, $\nu$, as shown in the example in **Fig. 3**.

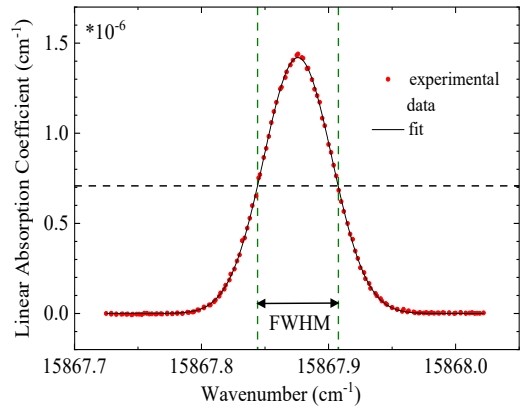

**Fig. 3.** Doppler-broadened absorption profile of $O(^3P_2)$ atoms. Red points are experimental data, and the black line is a Gaussian fit. The absorption away from the resonance (due to mirror losses and continuum absorption in the cavity) has been subtracted to set the baseline to zero in this figure.

The gas translational temperature can be calculated directly from the Doppler width, $\Delta\nu_D$ (full width at half maximum), using **Eq. 5**.

$$\frac{\Delta\nu_D}{\nu} = \frac{1}{c}\sqrt{\frac{8kT_g \ln 2}{m}} \qquad \text{Eq. 5}$$

where $T_g$ is the gas transitional temperature, $k$ is the Boltzmann constant, $m$ is particle mass and $\nu$ is the peak wavenumber.

The absolute density of oxygen atoms in the lowest spin-orbit state, $O(^3P_2)$ is obtained from the integrated area of this profile using **Eq. 3**. There have been no measurements of the integrated cross-section for the transition $O(^3P_2) \rightarrow O(^1D_2)$. Therefore, it must be calculated from the Einstein A coefficient, for which the only available values are from theoretical calculations. Taking the value given in the NIST data base of $5.65 \times 10^{-3}$ s$^{-1}$ (estimated uncertainty of ±7%) [39], and with $g_{up} = g_{low} = 2J+1 = 5$, gives a value of $\int \sigma_\nu^X d\nu = 2.977 \times 10^{-23}$ cm. At these temperatures, the other spin-orbit levels of the oxygen atom ground state, $^3P_1$ (at $E_1$=158.265 cm$^{-1}$) and $^3P_0$ (at $E_0$=226.977 cm$^{-1}$) will also be significantly populated. Peverall *et al*. [37] observed that the spin-orbit distribution was already in equilibrium with the gas translational temperature at 13 Pa, a conclusion that should be even more valid at the higher gas pressures of this study. Therefore, the total number density of O($^3P$) atoms can be derived from the O($^3P_2$) density using **Eq. 6**.

$$[O(^3P)] = [O(^3P_2)]\frac{1}{g_2}\sum_J g_J exp\left(\frac{-E_J}{kT_g}\right) \qquad \text{Eq. 6}$$

Finally, assuming that the oxygen atom density is zero in the narrow tubes between the mirrors and the active plasma (in which there is a strong flux of O$_2$ gas directed towards the plasma), the density in the active plasma region can be estimated from the line-averaged value through multiplying by the factor $L_{cavity} / L_{plasma}$ = 92 / 50.

### 2.2.3 Time-resolved measurements in modulated plasmas

For time-resolved measurements, the RF power is square-wave modulated, typically for 2 second on and 2 seconds off. During the pulsing, RF impedance matching is achieved by automatic RF frequency tuning, as explained above. Time-resolved CRDS measurements are made with the laser set to a fixed wavenumber. The ringdown events are not synchronized to the discharge pulsing but allowed to free-run (as for the measurements in the continuous discharge case). However, for each ringdown event the time delay relative to the most recent plasma-on transition is recorded. The data points are then sorted into time-bins (relative to the pulse cycle) and averaged, allowing the temporal profile to be reconstructed. Time-resolved measurements are made with the laser wavenumber set either to the peak of the Doppler profile (~15867.862 cm$^{-1}$, the "on-resonance" signal), or in the continuum region 0.13 cm$^{-1}$ to the low wavenumber side of the peak ("off-resonance" signal). In order to reduce the effect of slow drifts in $\tau_0$, the laser wavenumber is automatically switched between these two wavenumbers every 2 minutes, and the signal is averaged over several 10's of minutes. The absorption cross-section at the peak of the transition (where the on-resonance measurement is taken), depends on the Doppler width, and is therefore temperature-dependent. Considering the ratio of the peak of the Doppler profile to the integral of the absorption cross-section, we obtain for O($^3P_2$):

$$\sigma_{peak}^{O\,^3P_2}(T_g)(cm^2) = 9.875 \times 10^{-21}/\sqrt{T_g} \qquad \text{Eq. 7}$$

# 3. Results

## 3.1 Oxygen atom density and temperature in continuous discharges

Measurements were made in continuous plasmas over a range of pure $O_2$ gas pressures (67-800 Pa) and nominal powers (50 W – 900 W). Note that the measurements do not span the same power range at all pressures. At pressures up to 267 Pa, the maximum power is limited by the maximum current (20 amperes rms) allowed by the RF impedance probe. At higher pressures (533 and 800 Pa) and low powers, spatial-temporal instabilities occur below a certain power threshold, taking the form of moving plasma balls. Thus, it was only possible to make reliable measurements above this power threshold. In order to avoid artefacts caused by drifting surface behaviour, all measurements were made after the O atom density was observed to stabilise (typically after 10's of minutes).

The translational temperature is presented in **Fig. 4**. as a function of the RF power and the gas pressure. The gas temperature increases with power and pressure in nearly all cases, as would be expected. There is an exception at 133 Pa, where the temperature was seen to decrease with power over the range 270-330 W. This behaviour appears to correlate with a plasma mode change, which will be discussed below.

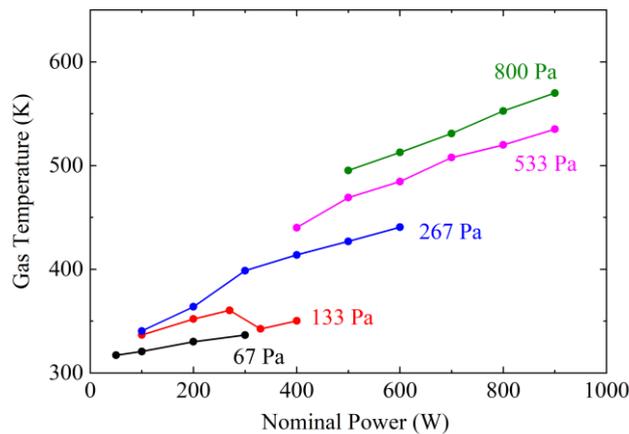

**Fig. 4.** Gas temperature in a continuous RF-CCP in pure oxygen as a function of O2 pressure and nominal RF power

The line-averaged oxygen atom density in the discharge is shown in **Fig. 5.** The O atom density trends are more complex. At 267 Pa and above, both the O density and the oxygen atom mole-fraction increase with power, but less than linearly. Indeed, at 267 Pa the atom density saturates at higher RF powers. At lower pressures the atom density first increases with pressure, passes through a maximum and then decreases.

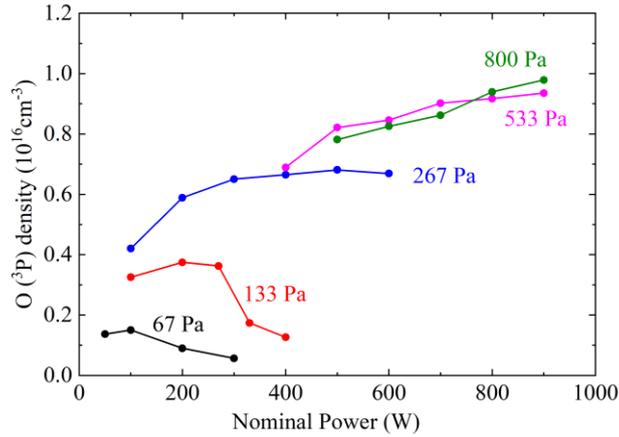

**Fig. 5.** Oxygen atom density in a continuous RF-CCP in pure oxygen as a function of $O_2$ pressure and nominal RF power.

To further understand these trends, it is useful to convert these data into the oxygen atom mole-fraction, using the measured gas temperature and the ideal gas law to calculate the total gas density, as plotted in **Fig. 6**. The mole-fraction reaches 15% at 267 Pa, 600 W. These high values are rather surprising in a reactor with metal surfaces, which are expected to effectively catalyse atom recombination. At higher pressures (267-800 Pa) the atom mole-fraction continually increases with RF power, but decreases significantly with gas pressure. At lower pressures the atom mole-fraction passes through a maximum with power, before a dramatic decrease, as was seen for the atom density. The mole-fraction at 67 Pa is smaller than at 133 Pa.

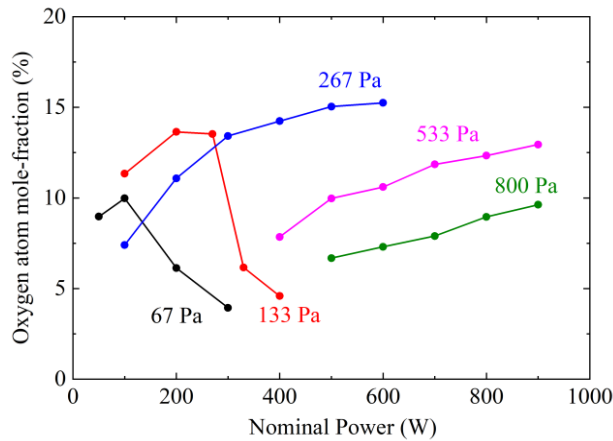

**Fig. 6.** O atom mole fraction in a continuous RF-CCP in pure oxygen as a function of $O_2$ pressure and nominal RF power.

The reasons for these trends will be discussed in more detail below, but here we will simply note that the steady-state O atom mole-fraction is the result of the balance between production (predominantly by electron impact dissociation of $O_2$) and loss by recombination processes (at surfaces and in the gas phase). The increasing dissociation with RF power (267 Pa and above) could be explained simply in terms of increasing electron density with RF power, causing increased electron-impact dissociation of $O_2$. In this hypothesis, the electron energy distribution function and the atom loss rates are assumed to be relatively constant at a given pressure. The significant decrease in atom mole-fraction with pressure can be attributed to changes in the EEDF, notably a significant decrease in the fraction of electrons with enough energy to cause dissociation.

The results at 133 and 67 Pa are more complex to explain. The substantial decreases at high power are unlikely to be caused by a decrease in electron density (which in general increases with power). Either the atom recombination rate (principally at the electrode surfaces at these pressures) increases dramatically with RF power, or the EEDF changes significantly, with a collapse in the fraction of high-energy electrons at high power.

Time-resolved CRDS measurements in the afterglow of time-modulated discharges allows the atom recombination processes to be probed. In the active plasma, electron impact induced dissociation is balanced by recombination reactions. In the afterglow only neutral processes remain, so probing the atom density decay provides information on the rates of these loss processes. In practice, the gas temperature is lower in the afterglow compared to the active plasma, which will reduce transport of atoms to the surfaces, as well as the rates of thermally-activated reactions. However, time-resolved kinetic studies in pulsed plasmas are the only experimental way to probe the atom loss rates in the plasma, despite this limitation.

## 3.2 Time-resolved CRDS in pulsed discharges

Examples of the time-resolved absorbance (both on-resonance with the oxygen atom transition, and off-resonance) are shown in **Fig 7** at 133 Pa and 270 W, with a duty cycle of 2 seconds on and 2 seconds off. Notably, the off-resonance signal (black curve in **Fig 7**) is not constant in time. This time-varying continuum absorption must originate from other species which absorb in this spectral region, which will be discussed below. However, to determine the time-behaviour of the absorption due to oxygen atoms, it suffices to subtract this baseline from the on-resonance data.

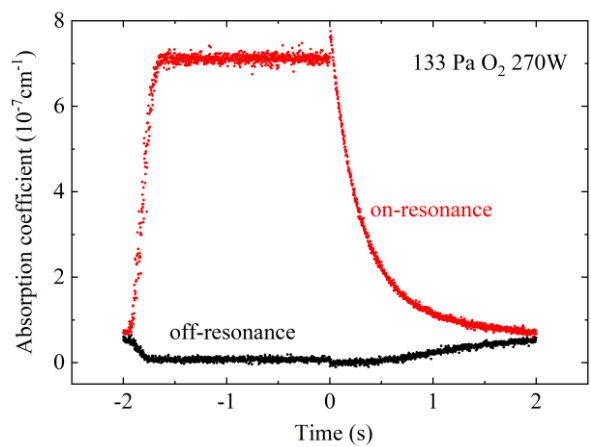

**Fig 7.** CRDS measurements at 133 Pa $O_2$, 270W power. RF power turned on at t = -2 s, and off at t = 0. On and off-resonance absorption.

### 3.2.1 Measuring oxygen atom kinetics

The time-resolved absorption due to oxygen atoms (with the continuum baseline absorption subtracted) is shown in **Fig. 8**, for conditions of 133 Pa, 270 W RF Power.

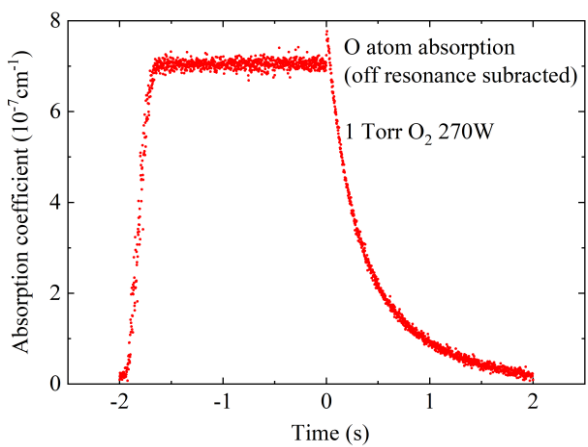

**Fig. 8.** O atom absorption (i.e., off-resonance signal subtracted from the on-resonance signal). 133 Pa (1 Torr) $O_2$, 270 W power. RF power turned on at $t$ = -2 s, and off at $t$ = 0.

After the discharge is ignited, the oxygen atom density stabilizes within a few hundred milliseconds. During the steady state discharge ($t$ = -1.7 to 0 s) the gas temperature is stable, and equal to that calculated from the steady-state Doppler width. When the RF power is stopped, the electrons cool very rapidly, and both the electrons and the ions are lost by diffusion to the walls within about 100 μs. Therefore, electron-induced dissociation of $O_2$ quickly stops, and we would expect the atom density to decay immediately. However, we observe that the absorption initially jumps up (by about 5%), before subsequently decaying. In the afterglow, the gas temperature (360 K in the active discharge) cools to the wall temperature (293 K). Considering the heat capacity (~900 Jkg$^{-1}$K$^{-1}$) and thermal conductivity (0.026 WmK$^{-1}$) of oxygen gas, this will occur with a time constant of about 5 ms at 133 Pa (and proportionately longer at higher pressures). The observed initial absorption increase can be explained by two mechanisms related to this gas cooling. Firstly, as the gas cools, the Doppler width decreases, and consequently the peak absorption cross-section increases. Secondly, during the steady-state discharge the gas temperature at the reactor mid-plane (where the measurements are made) will be higher than close to the electrodes, and consequently the gas density will be lower here. In the afterglow the temperature and density gradients will collapse, and convection will bring O atoms from the edge to the centre, leading to a spike in atom density before the density starts to decay.

After the gas has cooled (after about ~10 ms at 133 Pa) we can calculate the oxygen atom density assuming a constant peak cross-section of $\sigma_{peak}$ (O atom at 293 K)= $5.77\times10^{-23}$ cm$^2$ for the $^3P_2$ level (from **Eq. 7**), or $4.30\times10^{-23}$ cm$^2$ including all spin-orbit levels. The steady-state oxygen atom density determined in this way is consistent with the measurements in continuous plasmas as shown in **Fig. 5**.

## O atom density decay at lower pressures (up to 267 Pa)

The oxygen atom density decays in the afterglow are shown in **Fig. 9** for different gas pressures (up to 267 Pa) and RF powers. The oxygen atom density decays are normalised to aid comparison of the rates. The behaviour at higher pressures (533 and 800 Pa) is very different and will be discussed later.

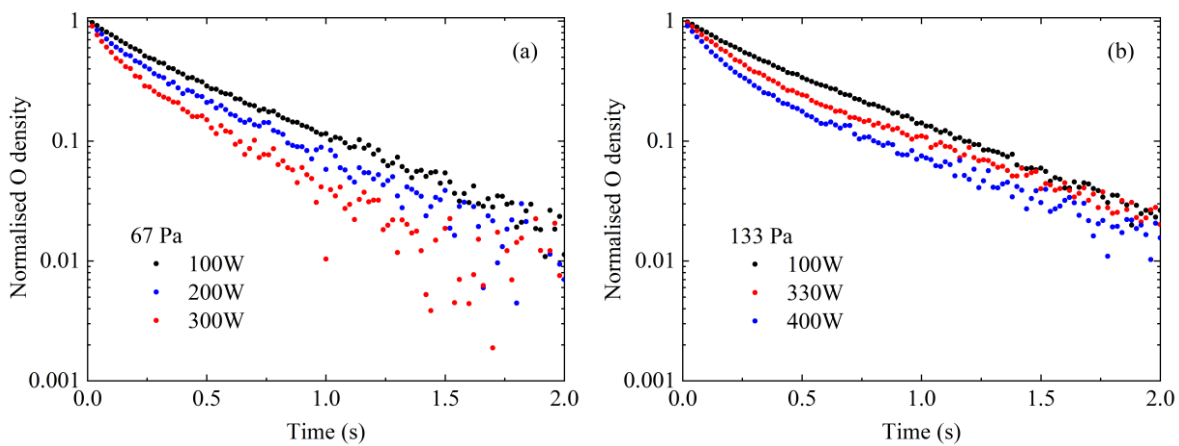

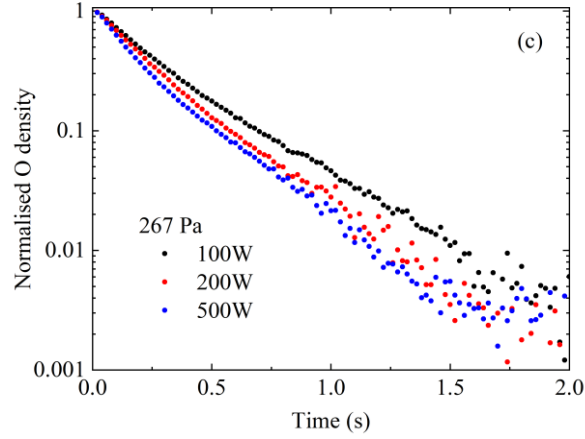

**Fig. 9.** Normalized time-resolved O density in afterglow at (a) 67 Pa, (b) 133 Pa, (c) 267 Pa.

At gas pressures up to 267 Pa, the observed decays are very slow, with some oxygen atoms still present (at $10^{-3}$ to $10^{-2}$ of the initial density) after 2 seconds. This is unexpected in a reactor with metal (aluminium) surfaces. Such slow decays indicate very small surface recombination coefficients, assuming that gas-phase processes are negligible (which is reasonable up to 133 Pa, as will be discussed below). These atom loss rates correspond to surface recombination coefficients of the order of only a few $10^{-4}$.

It can be clearly seen that the decay rates are faster at the start of the afterglow, and slow down with time, tending towards a constant rate at long times. Furthermore, the initial decay rates increase significantly with applied RF power. The data can be fitted well using the following function:

$$[O] = [O]_0 \times exp(-t(k_{final} + k_1 exp(-t/\tau)))$$  **Eq. 8**

In this expression, $k_{final}$ is the slow decay rate observed at long times, and $k_1$ represents an additional loss process present at the beginning of the afterglow, but whose effect decreases exponentially with time with a time-constant, $\tau$, which we refer to as the recovery time. Therefore, the initial decay rate (at the beginning of the afterglow, $t = 0$) is given by $k_{init} = (k_1+k_{final})$. An example fit (at 133 Pa pressure) is shown in **Fig. 10**.

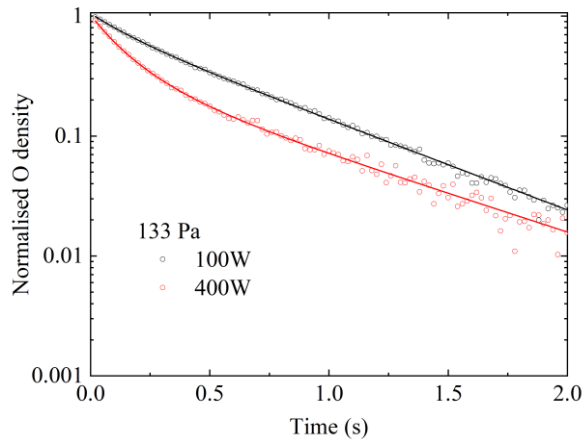

**Fig. 10.** Example of fitting by **Eq. 8** to measured O atom density decays in the afterglow at 133 Pa.

**Eq. 8** gave good fits to all of the data up to 267 Pa, and the results of the fits (the initial loss rate $k_{init}$, the final rate $k_{final}$, and recovery time, $\tau$, are summarized in **Fig. 11**. The final loss rates are significantly lower than the initial rates in all cases. The initial rate (**Fig. 11 (a)**) increases significantly with applied power, especially at lower pressure. This increase in initial oxygen atom loss rates (which is the rate closest to the loss rate during the active discharge) correlates well with the observed *decrease* in oxygen atom density as the power is increased (at 67 and 133 Pa). In contrast, the final decay rates only increase weakly with RF power.

Assuming that oxygen atoms are dominantly lost by recombination at the electrode surfaces, these observations can be explained by an increase in surface reactivity (creation of reactive sites on the surface) by energetic ion bombardment during the active plasma. These additional reactive sites then become passivated in the afterglow with a surface recovery time-constant, τ. This time-constant is of the order of about 0.5 seconds, but shows a complex dependence on pressure and power. Intriguingly, longer recovery times seem to be correlated with higher atomic oxygen density (**Fig. 5**).

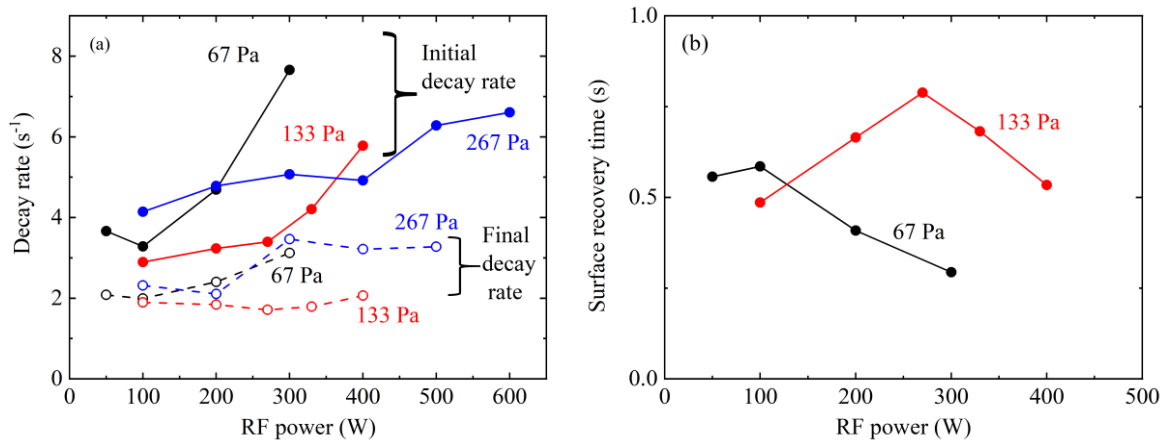

**Fig. 11.** Fitted parameter (a) initial rate, $k_{init}$, and final rate, $k_{final}$, and (b) Surface recovery time, $\tau$.

The highest initial loss rate (at 67 Pa 300 W) corresponds to a surface loss probability of $5.6 \times 10^{-4}$, whereas at 100 W it is only $2.4 \times 10^{-4}$, i.e., the surface loss rate is more than doubled as the power is increased. In the late afterglow at this pressure, it increases slightly with power, in the range $1.5\text{-}2.3 \times 10^{-4}$. The oxygen atom loss rates at 267 Pa are also shown in **Fig. 11**. However, at this pressure gas phase reactions start to play a significant role, and the results cannot be explained purely in terms of surface changes.

**O atom density decays at higher pressure (at 533 Pa and above)**

The afterglow behaviour of the O atom absorption (at resonance (15867.86 cm$^{-1}$) with the background absorption (at 15867.73 cm$^{-1}$) subtracted, normalised at t = 0) is shown in **Fig. 12** for pressures of 533 and 800 Pa.

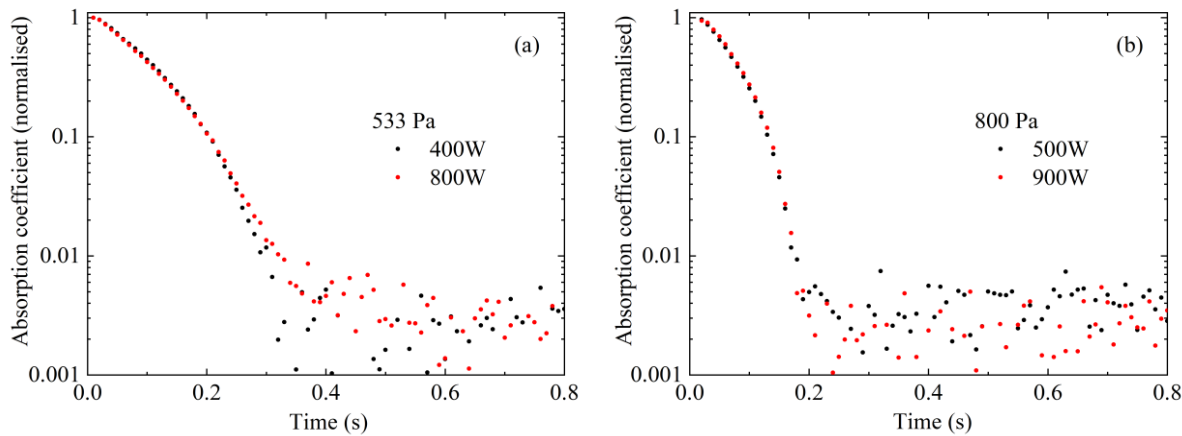

**Fig. 12.** Normalized time-resolved O density in afterglow at (a) 533 Pa and (b) 800 Pa.

The results obtained at these pressures differ dramatically from those at lower pressures in several respects. Firstly, the decays are significantly faster, falling to 1% of the initial values in 0.3 s at 533 Pa, and in 0.2 s at 800 Pa. Secondly, the decays start slowly and then accelerate with afterglow time. Moreover, the decay rates are not significantly affected by applied RF power. Finally, the absorption does not approach zero at long times; instead, it stabilizes at a nearly constant level—approximately a few $10^{-3}$ of the initial absorption—over the 2 s period following discharge termination. This absorption signal is unlikely to be due oxygen atoms, which could not persist in the afterglow for so long. It is more likely due to the presence of a stable molecule that absorbs photons in this spectral region, which has a bigger cross-section at 15867.86 cm$^{-1}$ than at 15867.73 cm$^{-1}$.

### 3.2.2　Continuum absorption: O$^-$ negative ions and ozone

The time-resolved absorption with the laser tuned off-resonance (wavenumber = 15867.73 cm$^{-1}$) at 133 Pa O$_2$, 270 W power, shown above in **Fig. 7**, is plotted on an expanded scale in **Fig. 13**.

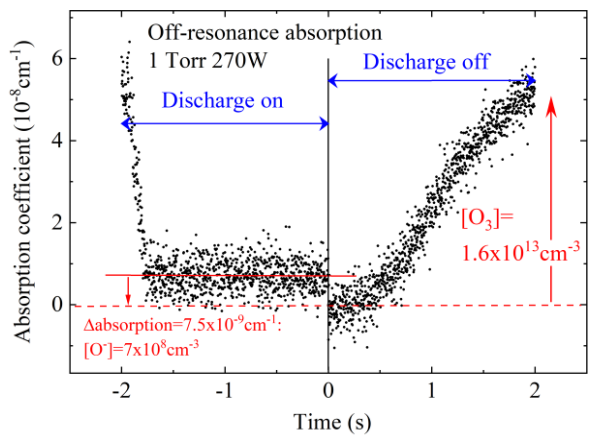

**Fig. 13.** Off-resonance (wavenumber = 15867.73 cm$^{-1}$) continuum absorption measured by CRDS at 133 Pa O$_2$, 270 W power, (expanded from **Fig. 7**).

We attribute this continuum absorption to two other species that are present in the system at different times: O$^-$ negative ions ($\sigma^{O-}$ = 5.8×10$^{-18}$ cm$^2$ at 630 nm, [40]) and ozone, O$_3$ ($\sigma^{O3}$ = 3.64×10$^{-21}$ cm$^2$ at 630 nm [41]).

The density of ozone in the active plasma is expected to be negligible, because it is destroyed quickly by electron impact. At longer times in the afterglow a slow rise in the continuum absorption is observed, which we attribute to ozone generation by gas-phase and surface reactions. In this example, the O$_3$ density increases up to 1.6×10$^{13}$ cm$^{-3}$ as shown in **Fig. 13**. It should be noted that this density is small compared to that of oxygen atoms under these conditions (3.6×10$^{15}$ cm$^{-3}$, i.e. about 2%).

Therefore, the background absorption in the active plasma can be attributed wholly to O$^-$ negative ions. When the plasma is first ignited, oxygen negative ions are created by fast dissociative attachment to O$_2$, and their density increases sharply, producing the observed fast rise in absorption at the beginning of the discharge pulse. However, these negative ions are efficiently destroyed by associative detachment reactions with O atoms and O$_2$($a^1\Delta_g$) molecules [42] as the density of these species builds up over the first few 10's ms of the discharge. After about 50 ms the O atom and O$_2$($a^1\Delta_g$) molecule densities reach their steady-state values, and the negative ion density falls to a steady state value. When the discharge is turned off, electrons diffuse to the reactor surfaces, and the electron density (and therefore the production of negative ions) falls to zero in a fraction of a millisecond. Negative ions are no longer produced, and the remaining ones are destroyed very rapidly (in a few μs, shorter than the time resolution of our measurements) by associative attachment reactions with O atoms and O$_2$($a^1\Delta_g$) molecules, leading to an almost instantaneous drop in the continuum absorption at discharge extinction (at $t$ = 0 s). Therefore, we can deduce the steady-state negative ion density from the amplitude of this step change (in this case, [O$^-$] = 2.37×10$^9$ cm$^{-3}$).

The O$^-$ negative ion density deduced from the absorption drop at the end of the discharge is presented in **Fig. 14** as a function of nominal power and O$_2$ pressure. Measurements were only possible up to 267 Pa. At higher pressures it was not possible to observe a clear drop in the absorption at plasma extinction, due both to lower negative ion densities and a higher (and varying faster with time) signal from ozone.

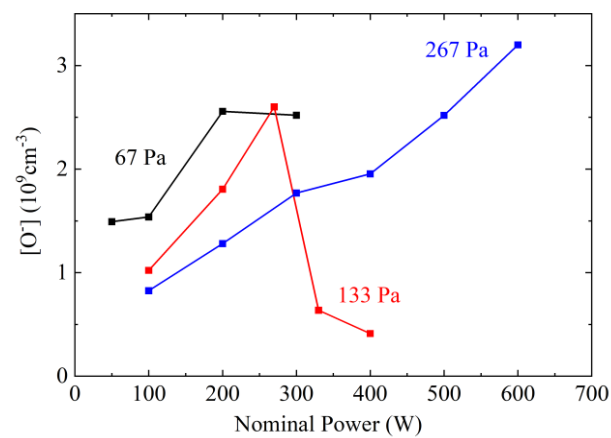

The negative ion density increases steadily with RF power at 267 Pa, whereas at 133 Pa it first increases, then drops dramatically after 270 W, the exact same conditions where we observed a strong decrease in the oxygen atom density. This further supports our hypothesis that there is discharge mode transition at this point. At 67 Pa the O$^-$ density first increases with power, then reaches a plateau between 200 and 300 W. We were not able to operate at higher RF power at this pressure due to the current limitation, so it is not possible to say if the density then decreases at higher powers.

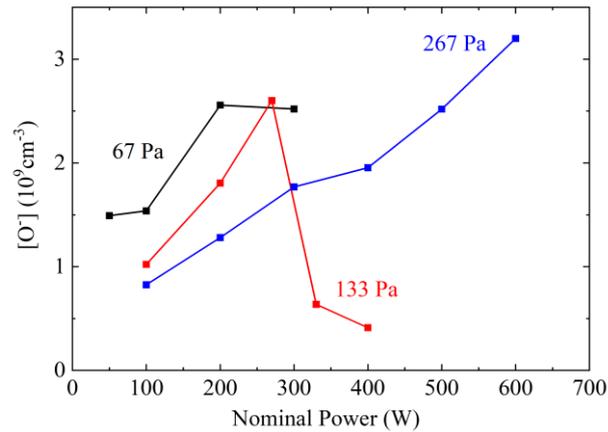

**Fig. 14.** O$^-$ negative ion density in the steady-state discharge, as a function of nominal RF power and gas pressure.

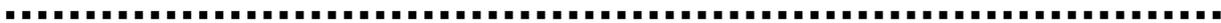

# 4. Discussion

Complete interpretation of the results presented here would require a full self-consistent plasma, chemical and thermal model, which is beyond the scope of the present work. Indeed, the aim of this article is to provide a comprehensive set of data, over a wide range of pressure and RF power, in order to test such models, which are under development by several groups. However, we can still draw useful qualitative conclusions about the mechanisms behind the observed trends.

## 4.1 Gas temperature

The gas temperature increases significantly with both injected RF power and with $O_2$ pressure (**Fig. 4**), with the exception of 133 Pa, where the temperature passes through a maximum with power (which will be discussed below). Direct momentum transfer from electrons to neutral particles through elastic scattering is inefficient at these pressures, due to the large mass difference between the colliding particles, and cannot account for the observed gas temperatures (Note that in atmospheric-pressure discharges this process can be significant, due to the much higher collision frequencies). However, two indirect processes can lead to significant gas heating [43]. ***Firstly, electron-impact dissociation*** occurs by the excitation from the ground state to dissociative excited states. Due to the longer equilibrium bond lengths in the upper, dissociative states, the Franck-Condon overlap favours excitation at energies above the dissociation limit. Consequently, the dissociating atoms depart with significant kinetic energy. ***Secondly, low-energy electrons can efficiently pump the vibrational states*** of the $O_2$ molecule via negative ion resonant states [44]. The vibrationally-excited molecules produced can then undergo efficient V-T transfer with oxygen atoms (a process which is much faster than V-T transfer between molecules) [45]. Annušová *et al.* [43] developed a global model including vibrational state kinetics, applied to an inductively-coupled plasma in pure $O_2$, at lower pressures than in this study (up to 13 Pa).

As the injected RF power is increased, the electron density increases, leading to an increase of the rate of both electron-impact dissociation and pumping of vibrational states, and thus the gas temperature increases. The effect of pressure is more complex. The rate of electron-neutral collisions will increase with pressure, leading to more gas heating through the mechanisms described above. The effect of pressure on the electron density is harder to predict. What is clear is that inelastic collisions will change the EEDF, shifting it towards lower energy at higher pressure. This will favour lower energy threshold processes such as vibrational excitation, and to some extent dissociation, over ionisation, so that a higher fraction of the injected power goes into gas heating.

## 4.2 Oxygen atom loss mechanisms

The oxygen atom density decays in the afterglow due to recombination reactions, both at the reactor surfaces and by three-body recombination reactions in the gas phase. To separate these effects, we must first discuss gas-phase recombination reactions, which have been studied for many decades by the atmospheric chemistry community.

### 4.2.1 Gas phase oxygen atom loss mechanisms

The principal gas-phase reactions of oxygen atoms occurring in a pure $O_2$ plasma are detailed in **Table 1**:

Table 1 Principal gas-phase O loss reactions

| | Reaction | Rate constant [reference] | Reaction Type |
|---|---|---|---|
| R1 | $O + O_2 + O_2 \rightarrow O_3 + O_2$ | $6 \times 10^{-34} \times (T_g/300)^{-2.37}$ cm$^6$s$^{-1}$ [46] | Three-body recombination |
| R2 | $O + O_2 + O \rightarrow O_3 + O$ | $2.1 \times 10^{-34} \times \exp(345/T_g)$ cm$^6$s$^{-1}$ [47,48] | |
| R3 | $O + O + O_2 \rightarrow O_2 + O_2$ | $3.81 \times 10^{-30} \times \exp(-170/T_g)/T_g$ cm$^6$s$^{-1}$ [49] | |
| R4 | $O + O_3 \rightarrow O_2 + O_2$ | $1.8 \times 10^{-11} \times \exp(-2300/T_g)$ cm$^3$s$^{-1}$ [50] | Bimolecular Exchange reaction |

Oxygen atoms are lost in the gas phase principally by three-body recombination, firstly with $O_2$ molecules to form ozone, $O_3$, and also by self-recombination to form $O_2$. The reaction with $O_2$ occurs through the formation of a vibrationally-excited intermediate, $O_3(v)$, which can dissociate back into O + $O_2$ unless it is stabilised by collision with a third body (principally $O_2$, via R1, but with some contribution from O, via R2). In the presence of high densities of oxygen atoms, the vibrationally-excited intermediate, $O_3(v)$ can also react with oxygen atoms to form $2O_2$, reducing the yield of ozone but increasing the loss of oxygen atoms. This is a complex subject, due to the large number of vibrational states involved, and several models have been proposed for this process [38].

Although the rates of mutual recombination reactions of atoms to form diatomic molecules are generally very small due to the rapid dissociation of the excited diatomic intermediate, in the case of oxygen it is more favoured due to the existence of several weakly-bound states (the A, A' and c states) close to the dissociation limit. Measurements of the rate of this process with $O_2$ as the third body are difficult to perform, due to competition from the reaction forming ozone. However, a number of measurements were made in the 1960's with $N_2$ or Ar as the third body, and have been combined with shock-tube measurements of the dissociation equilibrium of $O_2$ at high temperatures to allow an estimation of the rate constant used here [49]. Since the rate of this reaction is quadratic in the oxygen atom density, it will be of most importance when the atom density is high, i.e. at higher pressure and power, in the active discharge. It can also be significant at the beginning of the afterglow, but will become unimportant in the later afterglow.

Oxygen atoms can also be lost by reaction with ozone (R4). However, in the active discharge the density of $O_3$ is very low because it is destroyed quickly (by electron impact, as well as thermal decomposition). Even in the afterglow, the ozone density remains small (of the order a few $10^{13}$cm$^{-3}$) and the contribution of (R4) to oxygen atom loss can be ignored.

### 4.2.2 Surface oxygen atom loss mechanisms

The study of the recombination of atoms on surfaces has a long history, dating back to the work of Irving Langmuir [51,52], Cyril Hinshelwood and Eric Rideal in the 1920's and 30's. The recombination mechanisms are often classified as Langmuir-Hinshelwood (in which reaction takes place between two atoms adsorbed on the surface) and Langmuir-Rideal (in which an incident atom from the gas phase reacts directly with one absorbed on the surface).

In 1943 Smith [53] proposed a method to measure atom recombination coefficients, $\gamma$, from the atom density gradient along a closed tube with an atom flux supplied to the open end, and applied it to hydrogen atoms on borosilicate glass. Greaves and Linnett improved this technique [54], and determined a value of $\gamma = 1.6 \times 10^{-4}$ for oxygen atoms on silica at 300K, increasing to 0.01 at 900K [55]. They extended this work to a range of oxide surfaces [56], including $Al_2O_3$ ($\gamma = 0.002$) - although they believed that this value was an over-estimate due to the rough surface texture of the sintered material used. Larger values were observed for other surfaces, for instance 0.0052 on iron oxide and 0.043 on copper oxide. Kim and Boudart [57] used a development of Smith's technique to measure the recombination of O, N and H atoms on silica over a wide range of surface temperatures (194 to 1250K), and observed that the reactions were always first order in the incident flux of atoms, with a range of activation energies depending on the temperature range. They found that reactions occur at a small number of chemisorbed sites. For oxygen on silica they found values of $\gamma = 2 - 5 \times 10^{-4}$ over the temperature range 300-500K. Macko *et al*. [29] studied oxygen atom recombination in the afterglow of a DC discharge in a borosilicate glass tube over the temperature range 77-460K. They observed a similar behaviour with temperature as Kim and Boudart, although the values of $\gamma$ were higher by an order of magnitude. They explained their observations by a Langmuir-Rideal mechanism at higher temperatures, with the onset of a Langmuir-Hinshelwood mechanism at lower temperatures.

In most studies of atom recombination on surfaces the role of the plasma is not discussed; no distinction is made between measurements made on surfaces exposed only to atoms (downstream of a plasma), and surfaces exposed to an active plasma (in which the surface could be exposed to energetic photons and ions). However, Cartry *et al*. [58] studied oxygen recombination in a pulsed DC discharge in a borosilicate glass tube and observed a significant increase in the surface recombination rate with plasma exposure. Booth *et al*. [59] studied oxygen recombination in a partially-modulated DC glow discharge. They observed a strong correlation of the recombination coefficient with the gas temperature, which provided evidence for a Langmuir-Rideal mechanism. Furthermore, the observed dependence of the recombination coefficient on the magnitude of the incident atom flux indicated two distinct surface reaction sites, strongly-bound chemisorbed sites (constant in surface density) and a weaker-bound physiosorbed sites (whose surface density increases with incident atom flux). The latter mechanism leads to a recombination rate that is *quadratic* in incident atom flux. This effect was more clearly seen in time-resolved measurements of the atom decay in the afterglow [38], where the decays are initially significantly faster at the beginning of the afterglow due to the dominance of the quadratic term when the atom density is high. It should be noted that such a mechanism will only be significant for high oxygen atom densities, which were of the order $10^{15}$-$10^{16}$ cm$^{-3}$ in this study.

Gomez *et al*. [1] studied oxygen atom surface recombination in radiofrequency capacitive and inductive discharges on surfaces of several different compositions. In all cases they observed that the recombination coefficient increases significantly at lower pressures. They attributed this effect to the increasing influence of energetic ion bombardment of the surface at lower pressures. Booth *et al*. [59] also saw dramatic increases in the surface recombination on borosilicate glass in DC glow discharges for pressures below 67 Pa. The effect scaled linearly with ion flux, despite the modest estimated ion energies at the surface in this system (estimated up to 5 eV).

The group of Donnelly developed the spinning-wall technique to measure atom recombination mechanisms on surfaces, and presented results for oxygen atoms on anodised alumina surfaces. In the first study [60] they obtained values of 0.3-0.6. In a subsequent study [61] they reported a value of 0.06 for low atom fluxes, dropping to 0.04 at high atom fluxes. They attributed these differences to contamination of the spinning surface by silica sputtered from their ICP window. A comprehensive review of measurement techniques and observed values for oxygen atom recombination coefficients on various materials was made recently by Paul *et al*. [62]. The wide distribution of reported values highlights the importance of precise definition of all parameters (surface composition, history, temperature, incident flux of atoms, ions and (perhaps) photons) when specifying a recombination probability.

Finally, oxygen atoms can also combine with $O_2$ at surfaces to form ozone [38,63]. However, this pathway is minor compared to the production of $O_2$ and can be neglected as a major loss process for oxygen atoms.

### 4.2.3 Oxygen atom decays in the afterglow at lower pressures (up to 267 Pa)

At these lower pressures, we observe that the oxygen atom density initially decays faster at the beginning of the afterglow, then slows down with time. Furthermore, the initial decay rates increase significantly with the RF power at all pressures up to 267 Pa. The observed atom loss rate is the sum of gas-phase and surface processes. The surface loss rate can be estimated by subtracting the rate of the gas phase processes, which can be estimated from the rate constants given in **Table 1** along with the densities of O and $O_2$ and the gas temperature. The steady-state oxygen atom density and gas temperature are known from the CRDS measurements. The total gas density can be deduced from the ideal gas law, and therefore the density of $O_2$ molecules can be deduced from **Equation 9**:

$$[O_2] = \frac{p}{kT_g} - [O] \qquad \text{Eq. 9}$$

where $p$ is the gas pressure in Pa, $k$ is the Boltzmann constant, $[O]$ and $T_g$ are taken from the CRDS measurements shown in 错误!未找到引用源。 and **Fig. 5**. Note that a significant fraction of the $O_2$ molecules (up to around 10% [64]) can be in the metastable $a^1\Delta_g$ state. However, for the sake of simplicity, here we assume that this state has the same chemical reactivity as the ground state.

At the beginning of afterglow, the gas fractional composition will be comparable to that in the active plasma. In the steady state the gas temperature in the centre of the discharge (see 错误!未找到引用源。) is higher than the temperature of the walls (293K). In the afterglow the gas cools within 5-10 ms to equilibrium with the electrodes. The collapse of the axial centre-peaked temperature gradient (and corresponding hollow gas density profile) will cause gas to move from near the electrodes towards the centre. This temperature drop will also cause the pressure in the discharge volume to drop initially, an effect which becomes more significant at higher pressure. Nevertheless, due to the colder, denser gas near the walls, the amplitude of the effect is considerably smaller than the change in the central gas temperatures would suggest. The discharge volume is connected to a much larger manifold leading to the vacuum pump, such that in the afterglow some gas will be drawn back into the discharge volume until it approaches the discharge-on pressure. Time-resolved measurements of the pressure (obtained by monitoring the analogue output of the capacitance manometer connected directly to the discharge volume) showed that the pressure stabilises with a time-constant of about 200 ms. In the following we calculated the total gas density in the afterglow assuming a gas temperature of 293K and using the nominal gas pressure. The loss rates of oxygen atoms due to gas phase reactions estimated in this way are shown in **Fig. 15**. This figure shows both the initial rates for the major reactions, and the loss rate at long times (marked as "final"), which consists uniquely of the reaction with $O_2$ (since the O atom density is negligible at long times).

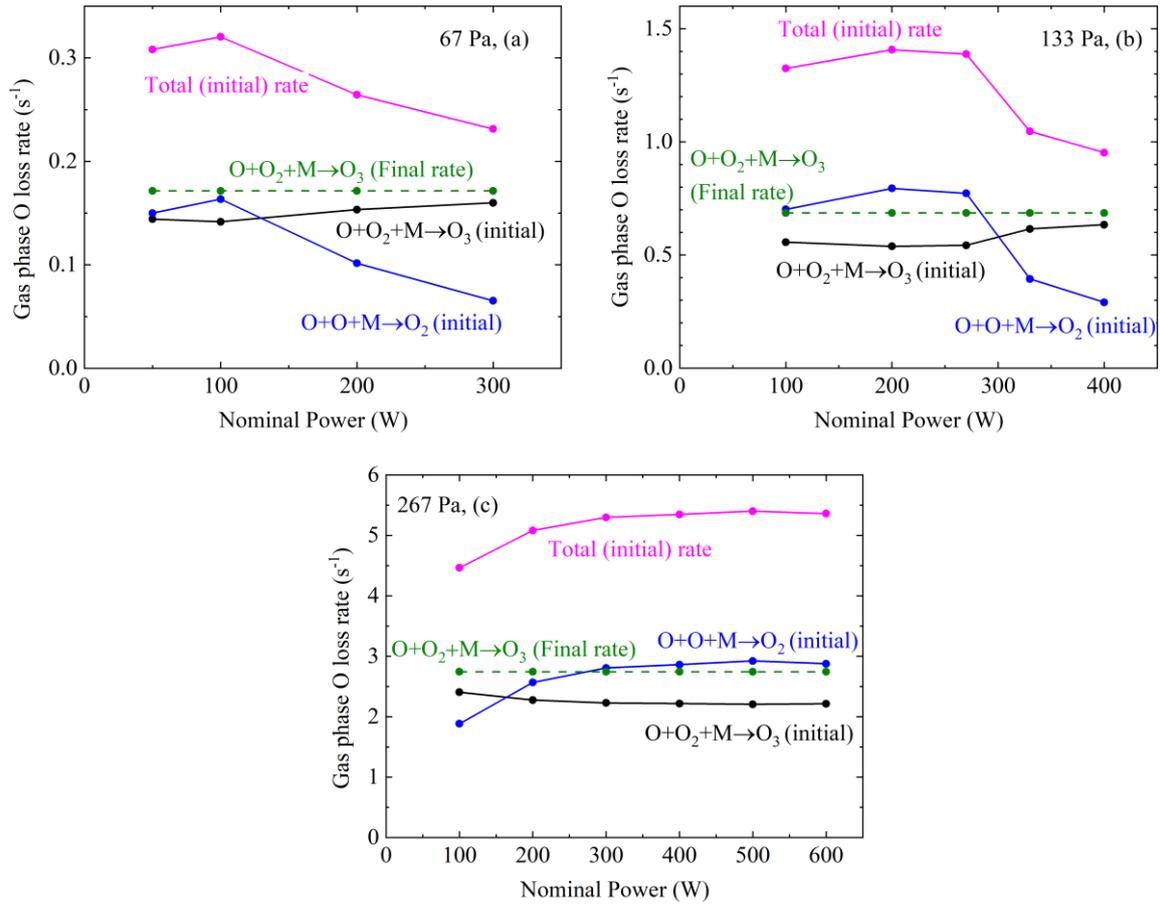

**Fig. 15.** Estimated gas phase O loss rates at (a) 67 Pa, (b) 133 Pa and (c) 267 Pa.

We can then estimate the oxygen atom surface loss rate by subtracting the estimated gas phase loss rates (**Fig. 15**) from the total observed loss rate (**Fig. 11**), as shown in **Fig. 16**.

At 67 Pa the rates of the gas-phase processes are negligible compared to the total loss rate, therefore recombination at the surface dominates. The dramatic increase of the oxygen atom loss rate with RF power (from about 3 to 8 s$^{-1}$) therefore indicates a substantial increase in surface reactivity.

At 133 Pa the rates of the gas-phase reactions are faster, but surface reactions still dominate the initial decay, and they again increase significantly with RF power. In the late afterglow the gas-phase reaction with $O_2$ to form ozone accounts for about 30% of the atom loss.

At 267 Pa the gas-phase reactions play a much bigger role; at low power they account for most of the initial loss rate. The increase in loss rate with power in this case is due to a combination of increased surface loss rate, along with an increase in rate of oxygen atom mutual recombination in the gas phase due to the higher O atom density. In the late afterglow, surface reactions become negligible, and the reaction with $O_2$ accounts for the majority of the observed loss.

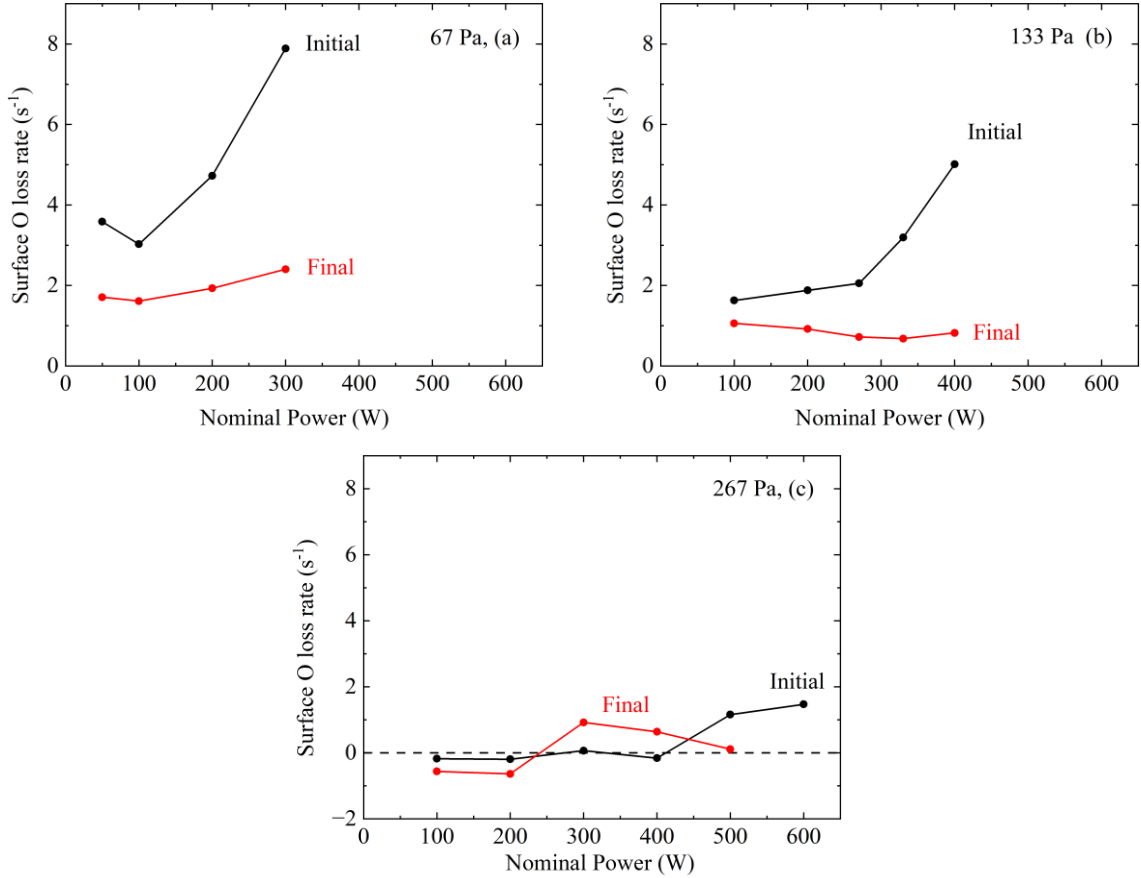

**Fig. 16.** Loss rates of oxygen atom due to surface reactions, deduced by subtracting the estimated rates of gas-phase reactions (**Fig. 15**) from the observed total decay rate in the afterglow (**Fig. 11**). (a) 67 Pa, (b) 133 Pa and (c) 267 Pa.

At these lower pressures, the energy of ions reaching the electrode surfaces will be significant, due to the relatively high sheath voltages and the lower collisionality of the sheaths, suggesting that this increased surface reactivity is induced by ion bombardment of the surface. This increased reactivity appears to decay in the afterglow with a time constant ($\tau_1$) of about 0.5 seconds, as presented in **Fig. 11 (b)**. This effect confirms the observations of Gomez *et al.* [1], who determined surface reaction coefficients from atom number density gradients close to surfaces, and saw that γ on aluminium increased from 0.01 at 67 Pa to 0.35 at 3.3 Pa.

A superficially similar effect of non-exponential (slowing) oxygen atom decays in the afterglow was observed in DC glow discharges in a glass tube in pure $O_2$ at similar gas pressures [38]. The non-exponential behaviour of the decays was observed to increase with the initial oxygen atom density, i.e. with both $O_2$ pressure and discharge current. These observations were attributed to the existence of two surface recombination mechanisms, comprising: 1) Langmuir-Rideal reactions with strongly-bound (chemisorbed) atoms, with a reaction rate, *a*, linear with the oxygen atom flux to the surface, and 2) Langmuir-Rideal reactions with weakly-bound (physiosorbed) atoms, with a reaction rate, *b*, quadratic with the oxygen atom flux to the surface. This model allowed the whole data set to be fitted with reasonably constant values of the parameters *a* and *b*. Measurements made while varying surface temperatures were coherent with this model. The corresponding surface recombination probabilities on borosilicate glass, γ, were of the order $1\times10^{-3}$ [59]. However, it is important to note that there are substantial differences between the two cases:

1) In the RF capacitive discharges studied here the sheath voltages reach several hundreds of volts, whereas in the case of the DC discharge, the borosilicate glass walls are at floating potential relative to the discharge (with a sheath potential of only ~15 V), and therefore the energy of ions reaching the surface remains quite low (estimated maximum 5 eV [38]).

2) In the current experiments the surfaces are aluminium metal (inevitably covered by a native oxide) compared to borosilicate glass in the DC glow experiments.

3) We observe here that the non-exponential nature of the decays increases strongly with the RF power, but decreases with pressure, and notably *does not correlate* with the initial oxygen atom density (which actually decreases at high RF power at 67 and 133 Pa). This is the opposite of what was observed in the DC discharge.

4) The effective surface recombination probabilities in the present case are significantly lower, of the order $1\times10^{-4}$.

Therefore, it appears that the mechanism leading to non-exponential decays is different in this case. It is consistent with ion bombardment-induced activation of the surface causing increased surface reactivity, an effect that appears to disappear with time in the afterglow. We observe that the initial decay rate increases with RF power (and therefore with the ion energy during the active discharge), but after about 0.5 seconds the decay rates become independent of the RF power.

### 4.2.4    Oxygen atom afterglow decays at higher pressures (533 and 800 Pa)

The results at 533 and 800 Pa are strikingly different to the lower-pressure results. Firstly, the decay rate starts slowly then *accelerates* with time in the afterglow. For instance, at 800 Pa the atom decay rate ($\frac{-1}{[O]}\frac{d[O]}{dt}$) increases from about 16 s$^{-1}$ at 40 ms in the afterglow to about 50 s$^{-1}$ at 130 ms. The decay rates are almost unaffected by the RF power (in the preceding discharge). Finally, the decay rates increase with gas pressure. At these pressures gas-phase reactions (with $O_2$ and O) dominate the loss at these pressures, and surface reactions are negligible. Under these conditions the sheaths are highly collisional, and ions reaching the surface will only have low energies, so that the RF power will have less effect on the surface state.

The increase in decay rate with time in the afterglow indicates that the densities of reaction partners in the gas phase is increasing with time. The gas temperature in the steady-state discharge increases with pressure (错误!未找到引用源。), therefore the effect of gas cooling and convection in the afterglow will also increase with pressure. As before, at the beginning of the afterglow the oxygen atom density actually rises, due to the collapse of the axial temperature and density gradients bringing O-rich gas to the centre from the regions near the walls. Then gas pressure drops, but then the total gas density increases as gas is drawn back into the discharge volume (with a time-constant about 200 ms). This increase in gas density in the later afterglow leads to a significant increase in the rate of three-body recombination reactions (quadratic with the gas density), which could explain the observed acceleration of the oxygen atom decay rate. We can estimate the loss rate of oxygen atoms by gas-phase reactions from the densities of O and $O_2$. Using the nominal gas pressure and a gas temperature of 293 K, the rate of loss of O atoms by recombination with $O_2$ to form ozone is 11 $s^{-1}$ at 533 Pa and 24 $s^{-1}$ at 800 Pa. This is significantly slower than the observed rate of decay (50 $s^{-1}$) at 130 ms, 800 Pa. A possible explanation could be that oxygen atoms are further reacting with the $O_3$ produced (reaction **R4** in **Table I**). However, to obtain an additional atom loss rate of 30 $s^{-1}$ would require an ozone density of $4\times10^{15}$ $cm^{-3}$. However, the off-resonance absorption under these conditions indicates an ozone density of only about $4\times10^{13}$ $cm^{-3}$, far too low to explain the observations. A possible mechanism could be the reaction of oxygen atoms with the vibrationally-excited $O_3(v)$ intermediate of reaction **R1**. The rate constant for this reaction could be much bigger than that for **R4**. However, a full quantitative explanation of these results is not possible at this time, and will require a model allowing for gas cooling and convection, as well as incorporating $O_3(v)$ kinetics.

Finally, we need to discuss the origin of the long-lived absorption baseline in **Fig. 12**. Firstly, we should note that the data shown in this figure represent the absorption at the peak of the oxygen line (15867.86 $cm^{-1}$) with the continuum absorption away from the peak (at 15867.73 $cm^{-1}$) subtracted. Therefore, it is not due to $O^-$ negative ions or $O_3$, whose continue do not show structure in this region. It is highly unlikely that this signal is due to long-lived oxygen atoms. It is possible that the absorption is due to the presence of nitrogen-oxygen compounds, created by the discharge due to the presence of air leaking into the chamber. Possible candidates are $NO_2$, which has a structured absorption spectrum in this region with a cross-section of about $1\times10^{-20}$ $cm^2$ [65], or the $NO_3$ radical with a cross-section of around $6.6\times10^{-18}$ $cm^2$ [66]. However, high-resolution spectra of these species are not available in this spectral region.

### 4.2.5 Trends in steady-state behaviour density with pressure and power, and possible discharge mode transition

We can now discuss the trends in the oxygen atom density and mole-fraction with RF power at different pressures, in the light of what we have learnt about loss mechanisms from the time-resolved afterglow measurements.

Firstly, at low pressure (67 and 133 Pa) we observed that the mole-fraction first increases with power, passes through a maximum then decreases strongly. At the same time, we observed that the atoms are lost predominantly by surface recombination, and that the rate of this process increases markedly with RF power, apparently due to the effect of energetic ion bombardment of the surfaces. The trend in mole-fraction with RF power could therefore be explained as the result of the competition between increased electron density causing increased $O_2$ dissociation, and increased surface recombination. However, a more quantitative examination suggests that this is not the whole story. At 133 Pa, increasing the RF power from 270 to 330 W causes the mole-fraction to drop from 13.5% to 6.2 %, while the surface reaction rate only increases from 2.1 to 3.2 $s^{-1}$. This suggests that there is also a drop in the $O_2$ dissociation rate, which could be caused by a significant change in the electron-energy distribution function, notably a collapse in the high-energy tail. Such a change could be induced by a so-called alpha-to-gamma mode change. Further evidence that this is occurring at this point is provided by the negative ion density (**Fig. 14**), which drops dramatically at exactly the same position (133 Pa, 270 → 330 W), as does the gas temperature (错误!未找到引用源。).

At 267 Pa and above, the oxygen atom mole-fractions increase progressively (but less than linearly) with RF power. As discussed above, the atom loss is dominated by gas-phase processes that vary little with RF power, and therefore these trends can be explained simply by increasing electron density (and $O_2$ dissociation by electron impact) as the power is increased, with no dramatic changes in the EEDF.

## 5. Conclusions

We have investigated the chemical processes occurring in an RF CCP in pure $O_2$ over the pressure range of 67 Pa to 800 Pa. The oxygen O atom density, translational temperature and loss rates in the afterglow are measured by CRDS. The loss mechanisms of oxygen atoms, including the plasma-surface interaction and the gas phase-surface competition, are analysed.

The oxygen atom density shows different behaviour with power at various pressures. Below 267 Pa, the oxygen atom density passes through a maximum with power. Time-resolved CRDS measurements in pulse-modulated plasmas indicate that the atoms are predominantly lost by surface recombination at low pressure, and the rates increase with RF power. This increased surface reactivity can be explained by increasing energetic ion bombardment. This effect partially explains why the mole-fraction passes through a maximum with power. We also see evidence of a discharge mode transition at 133 Pa, seen simultaneously in the $O_2$ dissociation rate, the gas temperature and the negative ion density.

In contrast, at 533 Pa and above, the oxygen atom density simply increases with power. Under these conditions the oxygen atom loss is dominated by gas-phase recombination reactions, and surface processes are negligible. This conclusion is supported by the time-resolved CRDS results, which show that the O atom loss rate in afterglow is independent of applied power. At these pressures the oxygen atom density increases simply due to the increased electron-impact dissociation. At these pressures the atom decay rate accelerates with time in the afterglow, which can be explained by a gas density increase caused by gas cooling and convection.